# On Lattices and the Dualities of Information Measures

David J. Galas[1,2*], Nikita Sakhanenko[1] and Benjamin Keller[1]

[1]Pacific Northwest Diabetes Research Institute

720 Broadway

Seattle, Washington 98122

USA

[2]Luxembourg Centre for Systems Biomedicine

University of Luxembourg

7, Avenue des Hauts-Fourneaux

L-4362 Esch-sur-Alzette

Luxembourg

* correspondence to:  David Galas, dgalas@pnri.org

## **Abstract**

Measures of dependence between and among variables, and measures of information content and shared information have become valuable tools of multi-variable data analysis. Information measures, like marginal entropies, mutual and multi-information, have a number of significant advantages over more standard statistical methods, including that they are inherently less sensitive to sampling limitations than conventional statistical estimates of probability densities. There are also interesting applications of these measures to the theory of complexity and to statistical mechanics. Their mathematical properties and relationships are therefore of interest at several levels.

There are a number of interesting relationships between common information measures, but perhaps none are more intriguing and as elegant as the duality relationships based on Möbius inversions. These inversions are, in turn, directly related to the lattices (posets) that describe these sets of variables and their multi-variable measures. In this paper we describe extensions of the duality previously noted by Bell to a wider range of measures, and show how the structure of the lattice determines some fundamental relationships of these functions. Our major result is a set of interlinked duality relations among marginal entropies, interaction information, and conditional interaction information, and some related observations. The implications of these results include a flexible range of alternative formulations of information-based measures, and a new set of sum rules that arise from path-independent sums on the lattice. Our motivation is to advance the fundamental integration of this set of ideas and relations, and to show explicitly the ways in which all these measures are interrelated through lattice properties. These ideas can be useful in constructing theories of complexity, descriptions of large scale stochastic processes and systems, and in devising algorithms and approximations for computations in this area that is becoming ever more important to multi-variable data analysis.

<u>Introduction</u>

Interaction information is an information measure for multiple variables that was introduced by McGill in 1954 [1]. It has been used effectively in a number of applications of information-based analysis, and has several interesting properties, including symmetry under permutation of variables. The interaction information for two variables is the same as mutual information, and conditional mutual information is the same as three variable interaction information (within a sign convention.) The interaction information for a set of variables or attributes, $v_n = \{X_1, X_2, X_3 \dots X_n\}$ obeys a recursion relation that parallels that for the joint entropy of a set of variables: $H(v_m) = H(v_{m-1}) - H(v_{m-1} \mid X_m)$

$$I(v_m) = I(v_{m-1}) - I(v_{m-1} \mid X_m)$$

As Bell first pointed out [2] there is an inherent duality between the marginal entropy and the interaction information based on Möbius inversion. Bell identified the source of this duality in the lattice associated with the variables. The duality is based on the partially ordered set of variables, ordered by inclusion, which corresponds to its power set lattice. We provide a direct proof of this symmetric property of the poset in the appendix, which is a variation on the usually stated form of the Möbius inversion.

<u>Möbius Inversion Dualities</u>

Consider a set of variables or attributes, $v_n = \{X_1, X_2, X_3 \dots X_n\}$ . We recall McGill [1] and adopt the sign convention of Bell [2] to define the interaction information[1] for the set of variables as

$$I(v) = \sum_{\tau \subseteq v} (-1)^{|\tau|+1} H(\tau)$$

$$(1)$$

where $\tau$ is any subset of $v$. Then, as Bell points out, and as can be easily shown, there is an elegant symmetry in the duality relation

---

[1] The recursion relation above can also be viewed as the definition of interaction information, which we will use later, but this definition is equivalent.

$$H(v) = \sum_{\tau \subseteq v} (-1)^{|\tau|+1} I(\tau)$$

(2)

The general relation, defined by equations 1 and 2, a form of symmetric Möbius inversion using the Möbius function of the power set lattice, can be easily demonstrated pictorially for three variables (figure 1.) If we use the Möbius function for the lattice shown[2] and identify the nodes as the marginal entropies for the variables indicated, we get the sum shown in equation 1 for three variables. Likewise, if we identify the nodes as the interaction informations for the variable subsets the sum yields equation 2. The key is that the poset, ordered by inclusion, can be used to map any pair of functions of subsets of variables, and the relation of the functions *H* and *I* in equation 1 then imply equation 2. The Möbius function defines the inclusion-exclusion relation for the poset ordered by inclusion.

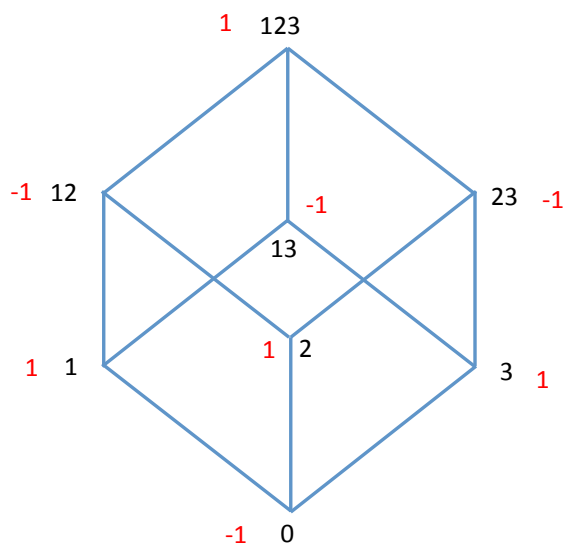

Figure 1. The power set lattice for three variables. The numbers in black are the variable subsets, while the Mobius function for this lattice is indicated in red.

---

[2] The convention we use here gives the function shown in figure 1, and allows the symmetry of the duality relations. The relation to the usual Möbius function of inclusion poset is indicated in the Appendix, and a direct proof is provided for this Möbius function.

To be specific, the *I-H* duality relations for *three variables* are these:

$$I(X_1, X_2, X_3) = H(X_1) + H(X_2) + H(X_3) - H(X_1, X_2) - H(X_1, X_3) - H(X_2, X_3) + H(X_1, X_2, X_3)$$
$$I(X_1, X_2) = H(X_1) + H(X_2) - H(X_1, X_2)$$
$$I(X_1) = H(X_1)$$

(3a)

and

$$H(X_1, X_2, X_3) = I(X_1) + I(X_2) + I(X_3) - I(X_1, X_2) - I(X_1, X_3) - I(X_2, X_3) + I(X_1, X_2, X_3)$$
$$H(X_1, X_2) = I(X_1) + I(X_2) - I(X_1, X_2)$$
$$H(X_1) = I(X_1)$$

(3b)

The well known and widely used multi-variable quantity, multi-information, or as it is often called, total correlation, is defined as

$$M(\{X_1 \ldots X_m\}) = \sum_{X_i} H(X_i) - H(\{X_1 \ldots X_m\})$$

(4)

This quantity also has the property that it goes to zero if the variables are independent, which is why it is often used. It is easy to show that if we limit the sum over subsets to those with two elements or more that the duality can be extended to include $I(v)$ and $M(v)$. We can use the duality expressed in equation 2 on the right side of equation 4 to get the expression for *M*, where the sum is over subsets $\tau$ that have 2 or more elements; $|\tau| > 1$

$$M(v) = \sum_{\tau \subseteq v} (-1)^{|\tau|} I(\tau)$$

(5a)

and the inversion yields

$$I(v) = \sum_{\tau \subseteq v} (-1)^{|\tau|} M(\tau)$$

(5b)

As a measure of its informsation content it is useful to determine how the interaction information changes when a new variable is added to the set. This can be indicated by the *difference* between the two interaction information measures for the expanded set and the original set of variables. Referring to the recursion relation for interaction information [1,2] we see that this differential interaction information is the same as the conditional interaction information (see equation 6.) This is a defining property of the interaction information. We note that interaction information could actually be defined as the information measure for a set of variables whose conditional on an added variable is identical to this *difference*. We indicate the subsets of $v_m$ that contain $X_m$ by $\{\tau_m \mid X_m \in \tau_m\}$. Then the differential interaction information, $\Delta$, is given by:

$$\Delta(v_{m-1}; X_m) \equiv I(v_m) - I(v_{m-1})$$
$$\Delta(v_{m-1}; X_m) = -I(v_{m-1} \mid X_m)$$

(6a)

We can then use equation 1, and equation 5b to obtain two interesting relations.

$$\Delta(v_{m-1}; X_m) = \sum_{\tau_m \subseteq v_m} (-1)^{|\tau_m|+1} H(\tau_m)$$
$$\Delta(v_{m-1}; X_m) = \sum_{\tau_m \subseteq v_m} (-1)^{|\tau_m|} M(\tau_m)$$

(6b)

where the second sum is over subsets with two or more elements. Recognizing that the subsets $\{\tau_m\}$ (containing $X_m$) map exactly (one-to-one) onto the subsets of $v_{m-1}$ (all subsets of variables) we can easily see that the lattice for $\Delta$ is simply a powerset lattice of one dimension smaller than the original set ($m$-1 versus $m$). Using this fact we can immediately derive a duality relation for $\Delta$. Rewriting the sum we have

$$\Delta(v_{m-1}; X_m) = H(X_m) + \sum_{\tau \subseteq v_{m-1}} (-1)^{|\tau|} H(\tau)$$

(7)

Thus, another duality relation follows directly

$$H(v_{m-1}; X_m) - H(X_m) = \sum_{\tau \subseteq v_{m-1}} (-1)^{|\tau|} \Delta(\tau, X_m) \tag{8}$$

The duality between $\Delta$ and $M$ can be similarly derived. Keep in mind that the differential interaction information is equivalent to conditional interaction information (equation 6). To explicitly illustrate the duality for three variables we again write out the full relations.

$$\Delta(X_1, X_2; X_3) - H(X_3) = H(X_1, X_2, X_3) - H(X_1, X_3) - H(X_2, X_3)$$
$$\Delta(X_1; X_3) = H(X_3) - H(X_1, X_3)$$
$$\Delta(X_2; X_3) = H(X_3) - H(X_2, X_3)$$
$$H(X_1, X_2, X_3) - H(X_3) = \Delta(X_1, X_2; X_3) - \Delta(X_1; X_3) - \Delta(X_2; X_3) \tag{9}$$

To illustrate this dimension reduction consider the four variable lattice (poset) for which we induce a dimension reduction by conditioning on variable four. This sublattice is shown in figure 2.

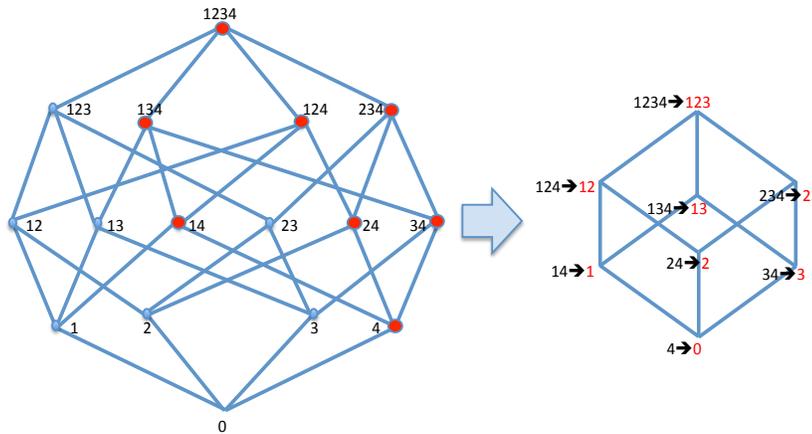

Figure 2. The four variable poset collapses to three dimensions when we condition on variable 4. The nodes of the three dimensional lattice are indicated in red, and the mapping back to the nodes for the three variable poset is shown on the lattice on the right.

What this says is that the conditioned quantities form a poset that has a lower dimension, three in this case, but that it maintains the power set structure. The more general relationship is illustrated in figure 3 below.

Another way of viewing the sub-lattice of everything above the node 4 in figure 2 is that it can be considered a "filter", and everything below the node 123 is an "ideal" and both are maximal (no other nontrivial filter or ideal is larger). (A "filter" is an upward closed set relative to the order, and an ideal is downward closed.) They are set complements, which makes them prime filters and ideals (of the 1234 lattice). These properties of the lattice reflect exactly the duality relations. To illustrate these properties further we specifically show the five variable lattice in figure 3, and show the equivalent set of sub-lattice relations.

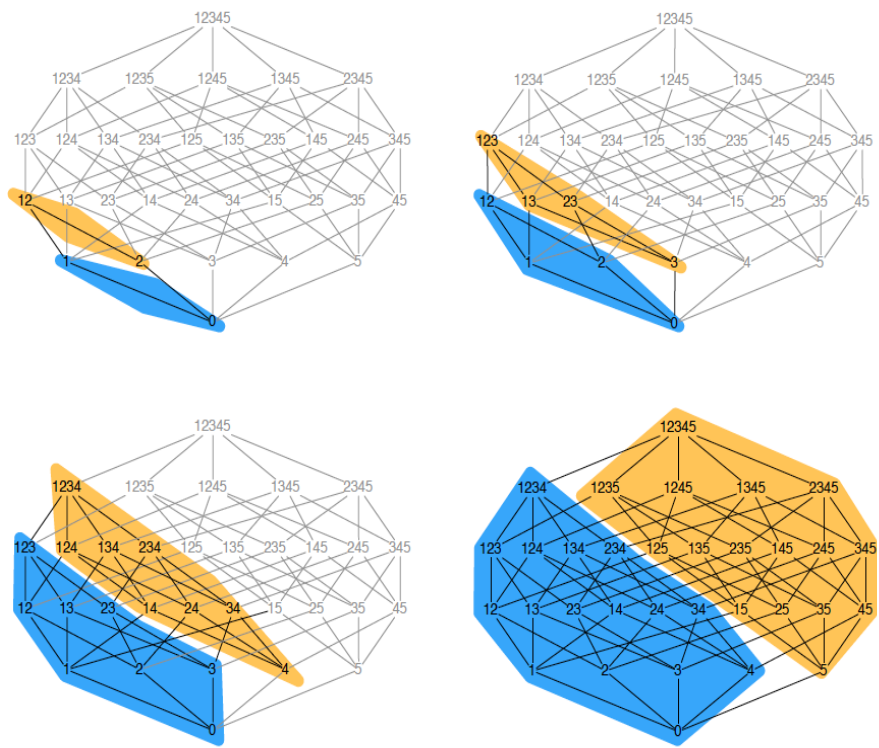

Figure 3. from upper left to right and top to bottom, the lattice for five variables is shown where the relationship illustrated in figure 2 is illustrated for 1, 2, 3 and 4 variables, showing 1, 2, 3 and 4 dimensional sublattices. The 3-D cube with the mapping shown in figure 2 is represented in the lower left drawing.

On the five-variable lattice the relationships are shown for 1, 2, 3 and 4 variables. The mapping shown in figure 2 is essentially the correspondence between the yellow and blue outlined sub-lattices (the 3-D lattice of figure 2 is shown on the lower left.)  The reduction in dimensionality corresponds to the conditioning on a variable.  Another way of thinking of conditional functions is that conditioning on a variable implies an embedding of a lattice of the original set of variables into a lattice of one higher dimension.  Some other connections are now evident.  It should be noted, for example, these symmetries are related to the symmetries of incidence algebras of functions on the lattice.

There is also a geometric analogy with these relationships.  The differential interaction information for $n$ variables, $\Delta(v_{n-1}, X_n)$ has the useful and interesting property that the symmetrized $\Delta$,  the product of all these $\Delta$'s obtained by permuting the variables (more precisely, picking the single asymmetric variable in all possible ways), is non-zero if only if the variables are collectively dependent [3].  If we think of the lattice as being embedded in a space of dimension $|v|$, and identify the spatial volume of the enclosed points of the $n$-1 dimensional solid with $\Delta$  (for example, the lattice in figure 1 becomes a real 3-D cube), then the structure consisting of these components with shared edges has an overall dimension of $n$ only if none of the $\Delta$'s is zero.   In the case of figure 1 we can identify the symmetrized $\Delta$ with the volume enclosed in the embedded space by the solid defined by the three edges, lengths equal to the three $\Delta$'s.  The volume of this structure is defined by the three edges if the length of an edge is the corresponding $\Delta$ for any dimension.  The structure collapses to zero volume if any one of the edges has length zero.   More generally we could embed the lattice in an $n$ dimensional space and identify the lengths of the edges with $\Delta$'s.  From any node of $m$ variables the volume subsumed by the set of m $\Delta$'s in the direction towards the meet of the lattice (towards fewer variable nodes) is identified with the symmetric $\Delta$.  The volume is the measure of dependence in this space and the Δ's are metrics. Thus, a non-zero Δ volume of a structure of $n$ dimensions means that the $n$ variables are collectively dependent.

The set of duality relations, which corresponds to the lattice structures of all these information measures, are shown in figure 4 below.

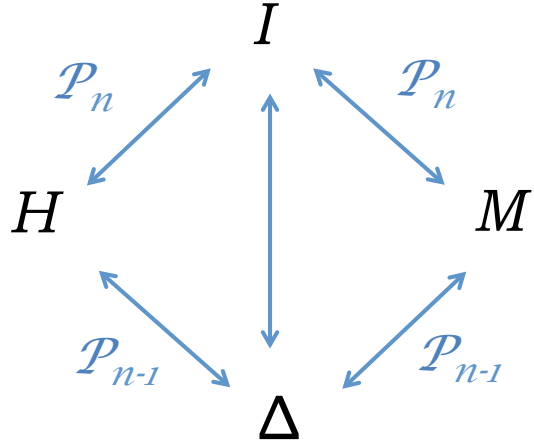

Figure 4. Four kinds of information measures ($I$, $M$, $H$ and $\Delta$) and their duality relationships, indicated by double-ended arrows. The dimensions of the power set lattices are indicated next to the arrows.

We give an example of the vertical duality, which we have not discussed explicitly, for three variables.

$$I(X_1, X_2, X_3) = \Delta(X_1, X_2; X_3) + \Delta(X_1; X_2) + H(X_1)$$
$$I(X_1, X_2, X_3) = \Delta(X_1, X_2; X_3) + \Delta(X_2; X_1) + H(X_2)$$

(10a)

or

$$I(X_1, X_2, X_3) - I(X_1) = \Delta(X_1, X_2; X_3) + \Delta(X_1; X_2)$$
$$I(X_1, X_2, X_3) - I(X_2) = \Delta(X_1, X_2; X_3) + \Delta(X_2; X_1)$$

(10b)

The complex of relationships and symmetries of these four information measures are rather simple, but they elucidate overlapping meanings and interpretations, which can otherwise be confusing. From the sets of dual equations it can be seen that additional dual relations can be derived.

<u>Conditional Measures and Path Independent Sums</u>

Using the lattice-variable relationships elucidated here we can easily extend the duality relations to measures conditioned on multiple variables. This is evident from the progressive dimension reduction in figure 2. Since conditioning on a variable implies an embedding of a lattice into a lattice of one higher dimension, each step down in dimension can be viewed as an added condition. It is easy to see this for interaction information directly from the recursion relation. Since

$$I(v_m) - I(v_{m-1}) = -I(v_{m-1} \mid X_m)$$

(11a)

if we condition on $X_k$ we can simply use this relation to show that

$$I(v_m \mid X_k) - I(v_{m-1} \mid X_k) = -I(v_{m-1} \mid X_m, X_k)$$
$$-I(v_{m-1} \mid X_m, X_k) = I(v_{m-1}, X_m, X_k) - I(v_{m-1})$$

(11b)

In this way the lower left lattice drawing in figure 3, for example, represents the adding of a conditioning on variable 4 to the blue sub-lattice in the lower right drawing (which itself represents the conditioning on variable 5.) Adding multiple conditional variables using the same procedure leads us to

$$-I(v_{m-1} \mid X_m, X_{m+1}....X_{m+k}) = I(v_{m-1}, X_m, X_{m+1}....X_{m+k}) - I(v_{m-1})$$

(12)

which is a direct reflection of the lattice structure and the dimension reduction discussed above.

If we consider the edges of the interaction information lattice (they have a direction so it's a directed graph) to have "weights" equal to the single variable conditional interaction information defined by the recursion relation of equations 6 and 11a, then a simple property emerges.   Since according to equation 11a the conditionals are simple differences between functions on adjacent lattice nodes, the sum of these conditionals on any path depends only on the end points (taking the direction into account by subtracting when traversing in the direction opposite to the lattice direction.)  That is, the sum of differences over any paths between the same two points is *path-independent*.  The sum of conditionals along any path between the same points is the same.  We can easily see this by adding conditionals as in equation 11 the intermediate nodes cancel and only the endpoints count.  This is what equation 12 suggests, and we can use this path independence to derive equalities between different sums and differences of conditionals.  Since these considerations are true for any function on the lattice it is simple to show that it is true of all of our information measure conditionals.   Furthermore, the difference between two unconnected sites in the lattice is the corresponding multi-conditional.

Note that any path in the directed graph that is the lattice is a fully ordered subset of the poset (called a chain.)  There are two kinds of sum rules that result from sums over chains.  First, there are those that result from the equivalence of sums over chains with the same end points, as mentioned above.  Second, there are those that result from those with the same final end point.  In this case the value of the lattice function at the starting nodes enters the equation.

All of the possible sum rules are available to define sets of relationships that may be convenient and useful in calculating and /or manipulating the information measures, and can also provide cross checks for some calculations.

<u>Conclusion</u>

We have shown that there is a set of duality relations among the information measures: interaction information, joint entropy, multi-information and differential interaction information. In addition, it is clear that a larger network of dual relations can be derived and added to this network if we include conditionals and higher differentials. The general theorem illustrated in the appendix points to fundamental nature of these dualities. They hinge not on the specific definitions or character of the information measure functions, but rather on the relationships of these functions to the posets ordered by inclusion.

The path independence of the set of conditioned functions, since it is defined in terms of differences, is the consequence of the basic lattice properties as well. Note also that the symmetric collective dependence measure for $m$ variables we defined previously [3] is simply the product of the edge weights of all the descending edges for the node for these variables in the lattice. This has an obvious geometic interpretation as an $m$-dimensional volume. All these relationships can serve to guide the use of the information measures together, help in the development of the theoretical characterization of complexity, and the algorithms and estimation methods needed for the computational analysis of multi-variable data. We can also see from this set of dualities that there may be much more to uncover in this complex of relationships. The information theory-based measures thus have a surprising richness and internal relatedness in addition to their practical value in data analysis.

**Appendix:**

**A proof of the symmetric inversion relation for posets ordered by inclusion**

Consider the set of variables $v = \{X_i\}$ for $i \in (1, n)$ and a pair of finite functions of any subsets of these variables, $\tau \subseteq v$, $f(\tau)$ and $g(\tau)$. The Möbius function is usually defined so that the relation between $f$ and $g$,

$$f(v) = \sum_{\tau \subseteq v} g(\tau)$$

can be inverted using the Mobius function

$$\mu(\tau, v) = (-1)^{|\tau| + |v|} \qquad g(v) = \sum_{\tau \subseteq v} \mu(\tau, v) f(\tau)$$

A symmetry of the respective relations can be achieved by factoring the above Möbius function. We then can state the following theorem defining the symmetric Möbius duality between functions $f$ and $h$. A short proof follows.

**Theorem**: If $f$ and $h$ are related by the relation $f(v) = \sum_{\tau \subseteq v} (-1)^{|\tau|} h(\tau)$ where the sum is over all subsets of $v$, then $h(v) = \sum_{\tau \subseteq v} (-1)^{|\tau|} f(\tau)$ .

Proof: We proceed by induction. If and only if the theorem is true substituting the two equations into one another we have

$$h(v) = \sum_{\tau \subseteq v} \sum_{\sigma \subseteq \tau} (-1)^{|\tau| + |\sigma|} h(\sigma) = \sum_{\sigma \subseteq \tau \subseteq v} (-1)^{|\tau| + |\sigma|} h(\sigma)$$

(A1)

It is easy to show by direct calculation when the set is two variables or three variables that all terms of the above sum over subsets of subsets cancel except the full set term. To illustrate, figure A1 shows the breakout of terms for three variables.

| 1,2,3 + | | | | | | | | **1,2,3** |
|---|---|---|---|---|---|---|---|---|
| 1,2 - | 1,2 + | | | | | | | **0** |
| 1,3 - | | 1,3 + | | | | | | **0** |
| 2,3 - | | | 2,3 + | | | | | **0** |
| 1 + | 1 - | 1 - | | 1 + | | | | **0** |
| 2 + | 2 - | | 2 - | | 2 + | | | **0** |
| 3 + | | 3 - | 3 - | | | 3 + | | **0** |
| 0 | | | | | | | | |

*Column headers (blue, above): 1,2,3 -   1,2 +   1,3 +   2,3 +   1 -   2 -   3 -   0   (rightmost: **Sums**)*

Figure A1.   This figure shows the variable terms (which variables, indicated by the numbers) and their signs.   The upper, blue numbers shows the $\tau$ subsets, and the matrix below shows the $\sigma$ sub- sets of the upper sub-sets.   They are aligned so that the sums can be read across.

We then assume the theorem to be true for the set $v$, of $n$ variables, and calculate the sum when we add a variable, $Y$, to the set:  $n+1$ variables.   All of the sub-sets of sub-sets are the same for the $n+1$ set but with the same added variable, $Y$,  except for the subset containing only $Y$.

The pattern of cancellation of terms is thus identical as for $v$, leaving only the full set term, and multiples of the subset terms containing only $Y$.  It is easy to see that these all cancel as well since there are  always an even number of these added together with alternating signs.  It's easy to see that the number of these terms is actually a power of 2 (the sum of all binomial coefficients, from zero to $n$.)   Thus, all terms cancel but the full set term, and the it must be true for arbitrary $n$.  Thus,  we have

$$h(\nu) = \sum_{\sigma \subseteq \tau \subseteq \nu} (-1)^{|\tau|+|\sigma|} h(\sigma) = h(\nu)$$

(A2)

which proves the theorem.